\documentclass[reprint, twocolumn, aps]{revtex4-2}
\usepackage{hyperref}
\usepackage{color}
\usepackage{multirow}
\usepackage{epsfig}
\usepackage{graphicx}
\usepackage{amsmath}
\usepackage{amssymb}
\usepackage{bm}
\usepackage{dcolumn}
\usepackage{braket}
\usepackage{bbold}
\usepackage{ulem}

\usepackage{graphicx}
\usepackage{dcolumn}
\usepackage{bm}
\usepackage{diagbox}
\usepackage{epsfig}
\usepackage{tabularx}
\usepackage{url}
\usepackage{chngcntr} 
\usepackage{textcomp}

\begin{document}

\preprint{APS/123-QED}

\title{X-ray magnetic circular dichroism of altermagnet  $\alpha$-Fe$_2$O$_3$ based on multiplet ligand-field theory using Wannier orbitals}

\author{Ruiwen Xie}
\affiliation{Institute of Materials Science, Technical University of Darmstadt, Darmstadt, Germany}
\author{Hamza Zerdoumi}
\affiliation{Institute of Materials Science, Technical University of Darmstadt, Darmstadt, Germany}
 
\author{Hongbin Zhang}%
\email{hzhang@tmm.tu-darmstadt.de}
\affiliation{Institute of Materials Science, Technical University of Darmstadt, Darmstadt, Germany}%


\date{\today}

\begin{abstract}
Hematite $\alpha$-Fe$_2$O$_3$ is a $g$-wave altermagnetic material, which has an easy-axis phase and easy-plane weak ferromagnetic phase below and above Morin transition temperature, respectively. The presence of these phases renders it a good candidate to study the characteristic spin splitting in altermagnets under the impacts of relativistic effect and finite temperature. In this regard, we have calculated the band structure of $\alpha$-Fe$_2$O$_3$ based on density functional theory (DFT) which also considers the Hubbard-U correction and spin-orbit coupling (SOC) effects. Additionally, the charge self-consistent DFT + dynamical mean-field theory (DMFT) calculations have been performed at finite temperatures. It is found that the altermagnetic spin splitting in $\alpha$-Fe$_2$O$_3$ preserves with either SOC or temperature effect taken into account. Furthermore, we present a numerical simulation of the x-ray magnetic circular dichroism (XMCD) of the L$_{2,3}$ edge of Fe using a combination of DFT with multiplet ligand-field theory (MLFT). In terms of the different N\'eel vectors present in $\alpha$-Fe$_2$O$_3$, we calculate the x-ray absorption spectroscopy (XAS) of the L$_{2,3}$ edge of Fe in the form of conductivity tensor and analyze the XMCD response from a perspective of symmetry. A characteristic XMCD line shape is expected when the N\'eel vector is along [010] direction (magnetic point group $2^\prime/m^\prime$) and the light propagation vector is perpendicular to the N\'eel vector, which can be further distinguished from the XMCD response originated from weak ferromagnetism with the light propagation vector parallel to the N\'eel vector.

\end{abstract}

\maketitle



\section{\label{sec:intro}Introduction}
Recently altermagnets have attracted a great amount of attention due to their intriguing properties, including anomalous Hall effect (AHE),\cite{vsmejkal2020crystal,vsmejkal2022anomalous,gonzalez2023spontaneous,naka2022anomalous} linear magneto-optical effects,\cite{samanta2020crystal,hariki2024x,hariki2024Ru,sasabe2023ferroic} and other phenomena such as spin Hall effects.~\cite{hu2025spin,watanabe2024symmetry} 
The altermagnets are characterized by the macroscopic time-reversal symmetry breaking, giving rise to finite non-relativistic momentum-dependent spin splitting. 
As a matter of fact, the macroscopic time-reversal symmetry breaking in some special types of antiferromagnets has been proposed by Dzyaloshinskii in 1950s,\cite{dzyaloshinsky1958thermodynamic} highlighting their capability of hosting weak ferromagnetism, piezomagnetism,\cite{dzialoshinskii1958problem} and magnetoelectricity.\cite{dzyaloshinskii1959magneto} 
Such an aspect was revisited alongside Turov’s phenomenological classification of these unconventional antiferromagnets as either ``centrosymmetric” or ``centroantisymmetric,”\cite{turov1990kinetic} 
and it was argued that a deeper analogy exists between certain classes of centrosymmetric antiferromagnets with broken time-reversal symmetry and ferromagnets. 
Namely, such antiferromagnets can be described using only a single magnetic site in the unit cell.\cite{solovyev2025hidden}
This analogy on the other hand raises a challenge in distinguishing the effects originated from altermagnetic and weak-ferromagnetic contributions. 

The x-ray magnetic circular dichroism (XMCD) technique demonstrates an advantage in this aspect that the valence spin-orbit coupling (SOC) plays a minor role in determining the spectral feature of XMCD.\cite{hariki2024x,kunevs2025neel} 
XMCD is dominated by the SOC of core states, which can be readily distinguished from weak ferromagnetism and other valence band effects.\cite{hariki2025separating} 
Previously, XMCD in $\alpha$-MnTe,~\cite{hariki2024x} rutiles RuO$_2$~\cite{hariki2024Ru}, MnF$_2$~\cite{hariki2024determination}, and NiF$_2$~\cite{hariki2025separating}  has been studied and discussed in terms of symmetry. 
For $\alpha$-MnTe, both theoretical and experimental studies have confirmed a characteristic XMCD line shape when the compensated moments align perpendicular to the light propagation vector, which is unconventional for XMCD measurements, as the light propagation vector typically lies parallel to the magnetic moments.\cite{hariki2024x}
Later, based on group theory, Kuneš derived a general relationship between the Hall vector and the orientation of the N\'eel vector in altermagnets under the free valence spin (FVS) approximation. 
It was found that XMCD may be present within the FVS approximation, emerges only when the multipolar core-valence exchange terms and valence SOC are included, or be fully forbidden, depending on the system symmetry.\cite{kunevs2025neel} 

Out of many possible altermagnets which can be straightforwardly identified by spin group theory,\cite{wan2025high} hematite $\alpha$-Fe$_2$O$_3$ is of particular interest, which experiences the Morin transition at $T_{\mathrm{M}} \approx 250$ K,\cite{morin1950magnetic} below which it demonstrates an easy-axis phase while at high temperature it hosts an easy-plane phase with a weak ferromagnetic moment arising from the Dzyaloshinskii-Moriya interaction.\cite{dzyaloshinsky1958thermodynamic,moriya1960anisotropic} 
In addition, the high-temperature weak ferromagnetic phase is stable up to the N\'eel temperature $T_{\mathrm{N}} \approx 955 \mathrm{~K}$.\cite{morin1950magnetic} 
Recently, $\alpha$-Fe$_2$O$_3$ was also identified as a $g$-wave altermagnet,\cite{vsmejkal2022beyond} which shows spin (chirality) splitting in the electronic (magnon) band structure.\cite{chen2025unconventional,hoyer2025altermagnetic} 
These magnetic properties make hematite a promising candidate for spintronic applications.\cite{baltz2018antiferromagnetic, lebrun2018tunable, ross2019propagation}

The Morin transition of $\alpha$-Fe$_2$O$3$ makes it well suited for studying the symmetry dependence of the XMCD response and for differentiating altermagnetism from weak ferromagnetism. 
In this work, we carry out $ab$-$initio$ calculations to investigate the electronic structure and XMCD for $\alpha$-Fe$_2$O$_3$, namely, the former obtained using the self-consistent density functional theory plus dynamical mean-field theory (DFT+DMFT) approach and the latter by constructing and solving the multiplet ligand-field model~\cite{haverkort2012multiplet}, as summarized in Sec.~\ref{sec:method} B. 
It is found that the non-relativistic spin splitting is quite robust against electron-electron interaction at finite temperature.
Detailed symmetry analysis has been conducted for XMCD, where the different features of XMCD originated from altermagnetism and weak ferromagnetism are demonstrated.
 
\section{\label{sec:method}Methodology}
\subsection{Band structure}
The band structure of $\alpha$-Fe$_2$O$_3$ was calculated using DFT+DMFT in a charge self-consistent manner,\cite{blaha2020wien2k,haule2010dynamical} in which the nominal double counting was adopted.\cite{haule2010dynamical}
The continuous-time quantum Monte Carlo (CTQMC) method was utilized to solve the quantum impurity problem\cite{haule2007quantum} with 600 million Monte Carlo steps. 
The charge density of the system at each self-consistent loop was converged to the accuracy of 10$^{-5}$. The Coulomb interaction $U$ and Hund's coupling $J$ were set to 7.0 and 0.8 eV, respectively, with the Coulomb repulsion described by the density-density form.  For comparison, the band structures were also calculated using the primitive cell of $\alpha$-Fe$_2$O$_3$ based on DFT and DFT+U approaches, as implemented in the Vienna ab-initio simulation package (VASP).\cite{kresse1996efficient}
An energy cutoff of 520 eV and a k-mesh of $10\times10\times10$ were adopted with the energy convergence criteria being 10$^{-8}$ eV. 
In terms of the DFT+U calculation, the method introduced by Dudarev\cite{dudarev1998electron} was used with $U=3.5$ eV. It will be shown later that this value of $U$ gives similar band gap as that obtained by the DFT+DMFT calculations.

\subsection{XMCD}
The XMCD spectra of $\alpha$-Fe$_2$O$_3$ were obtained using DFT combined with multiplet ligand-field theory (DFT + MLFT).\cite{haverkort2012multiplet} 
After self-consistent DFT calculations using the full-potential local-orbital (FPLO) package\cite{koepernik1999full,opahle1999full} with a k-mesh of 30$\times$30$\times$30, the symmetry-conserving maximally projected Wannier functions (SCMPWF)\cite{koepernik2023symmetry} were constructed for Fe 3$d$ and O 2$p$ orbitals. The constructed Wannier functions well represent the band structure obtained directly from DFT (see Fig.~\ref{fig:wann}). 
It should be noted that for Fe1 site (cf. Fig.~\ref{fig:band}(a)), the local axes in Cartesian coordinates $a^{\prime}$([100]), $b^{\prime}$([010]), and $c^{\prime}$([001]) are adopted for Wannier function construction. 
Following the symmetry relation, e.g., for Fe2 site which is related to the Fe1 site by a C$_{2y}$ rotation, the local axes of Fe2 site are $-a^{\prime}$, $b^{\prime}$ and $-c^{\prime}$. The Fe3 and Fe4 sites are connected with Fe2 and Fe1 sites, respectively, through inversion relation. Correspondingly, the local axes of Fe3 and Fe4 are kept the same as Fe2 and Fe1, respectively. 

The tight-binding model representing DFT Hamiltonian was set up considering a cluster with radius of around 2.2 \AA.  The Slater integrals were computed from DFT radial wavefunctions and listed in Tab.~\ref{tab:slater}. Additionally, in MLFT calculations, the double-counting term was carefully corrected as demonstrated in Ref.~\onlinecite{haverkort2012multiplet}. In this regard, we constructed an atomic Hamiltonian which addresses also the ligand-metal charge transfer effect
\begin{equation}
	\label{eq:ctm}
	H_{\mathrm{CTM}}=H_U^{d d}+H_U^{pd}+H_{l\cdot\mathrm{s}}^d+H_{l\cdot\mathrm{s}}^p+H_o^d+H_o^p+H_o^L+H_{\mathrm{hyb}}^{d L}+H_{\mathrm{ex}},
\end{equation}
in which $H_U^{d d}$ and $H_U^{pd}$ denote the Coulomb interactions between the 3$d$ electrons of Fe, and the Fe 2$p$ and 3$d$ electrons, respectively. $H_{l\cdot\mathrm{s}}^d$ and $H_{l\cdot\mathrm{s}}^p$ represent the spin-orbital coupling (SOC) term for Fe 3$d$ and 2$p$ orbitals. The on-site Hamiltonians corresponding to Fe 3$d$, 2$p$ and ligand orbitals are denoted by $H_o^d$, $H_o^p$ and $H_o^L$, respectively. $H_{\mathrm{hyb}}^{d L}$ defines the hybridization interaction between Fe 3$d$ and the ligand orbitals, and $H_{\mathrm{ex}}$ is the Weiss magnetic field acting on Fe, which represents the magnetic order of the Fe atom. In the calculation of XMCD, the Weiss field is set to 0.01 eV, which results in a spin moment of Fe around 4.60 $\mu_B$. The spectroscopy can then be calculated in accordance with Fermi’s golden rule using the equation
below:
\begin{equation}
	\label{eq:xmcd_g}
	G_\pm(\omega)=\langle\psi| T_\pm^{\dagger} \frac{1}{\left(\omega+\mathrm{i} \Gamma / 2+E_i-H_{C T M}\right)} T_\pm|\psi\rangle.
\end{equation}
The difference between the responses brought by right ($T_{+}$) and left ($T_{-}$) circularly polarized light gives rise to the XMCD signal. 
It should be noted that the element-specific XMCD of the crystal consists of contributions from all atoms. Unlike conventional collinear antiferromagnets, altermagnets can exhibit XMCD because their magnetic symmetry permits an axial vector response despite zero net magnetization.\cite{hariki2024x}

\section{\label{sec:result}Results and Discussion}
\subsection{\label{sec:band}Altermagnetic splitting in electronic band structure}
\begin{figure*}[th!]
	\centering
	\includegraphics[width=1.0\linewidth]{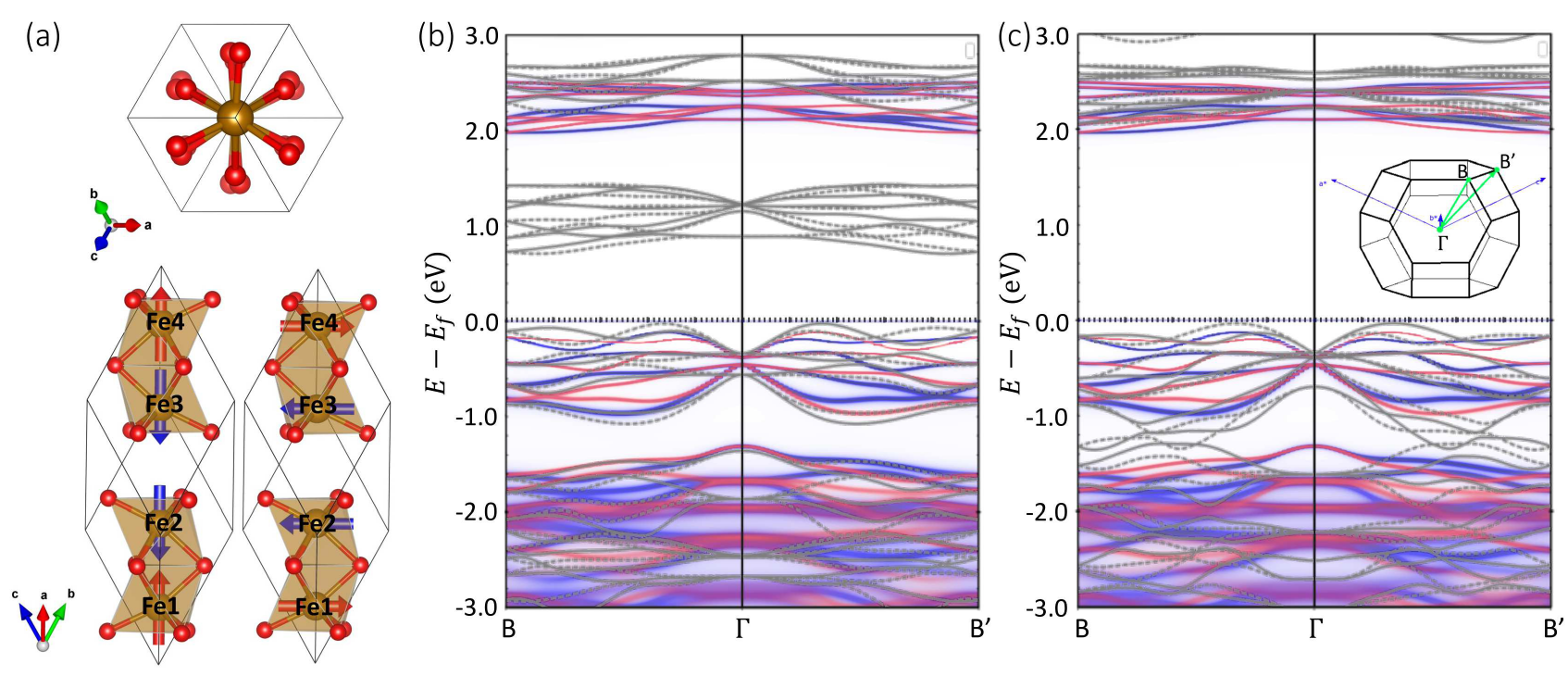}
	\caption{\label{fig:band}(a)Top view (upper panel) and view along [001] direction (in Cartesian coordinates, lower panel) of the primitive cell of $\alpha$-Fe$_2$O$_3$. The directions of the magnetic moment below (left lower) and above (right lower) Morin transition are illustrated, which are corresponding to N\'{e}el vectors along [001] and [010] directions, respectively. (b)Band structure calculated by DFT+DMFT method at 116 K (coloured contour with red and blue colours representing spin-up and spin-down channels, respectively), and by the DFT method (grey line with solid and dashed line representing spin-up and spin-down channels, respectively). (c) The same DFT+DMFT-calculated band structure as in (b), but compared to the DFT+U results. The inset marks the selected k-path in the Brillouin zone.}
\end{figure*}

Hematite $\alpha$-Fe$_2$O$_3$ has a space group $R\overline{3}c$ (No. 167) in the Hermann Mauguin notation, which is centrosymmetric and has a rhombohedral primitive cell (see Fig.~\ref{fig:band}(a)). The lattice parameters used in this work are $a=5.19$ \AA~and $c=13.77$ \AA. 
The chiral splitting of magnons caused by altermagnetic ordering has been predicted using Heisenberg exchange parameters evaluated by the first-principles calculations.\cite{vsmejkal2023chiral}
Later a four-sublattice spin-wave theory was developed considering the two different magnetic configurations undergoing the Morin transition and has shown that SOC does not obscure the altermagnetic features. 
As shown in Fig.~\ref{fig:band}(b), the altermagnetic splitting in the band structure can be captured by DFT+DMFT calculation at around 116 K. 
The band gap of approximately 2.1 eV is derived, which aligns well with the reported values by experiment.\cite{tahir2023investigation}
The bare DFT calculation underestimates the band gap but the band characters, {\it e.g.}, the valence bands below $E_f$, are reasonably aligned with the DFT+DMFT results. 
On the other hand, adding Hubbard $U$ correction results in a larger gap opening, giving rise to a comparable band gap as derived from the DFT+DMFT results (see Fig.~\ref{fig:band}(c)). 
However, the valence band features are also largely altered, {\it e.g.}, the location of the valence band maximum and the band splittings, which can be attributed to the many-body interactions considered in DMFT calculations.

The band structure demonstrated in Fig.~\ref{fig:band} does not include the SOC effect. In order to examine if the altermagnetic splitting remains intact, we further performed DFT+U+SOC calculations for the two magnetic configurations with N\'{e}el vectors ($\hat{L}$) directed along [001] and [010] directions, respectively. 
As can be seen from Fig.~\ref{fig:band_soc} in Appendix, the inclusion of SOC has a negligible effect on the band structure and the altermagnetic splitting exists for both $\hat{L}=[001]$ and $\hat{L}=[010]$. 
In addition, the robustness of the altermagnetic splitting with respect to temperature is investigated. For $T=116$ K, the local magnetic moment of Fe obtained from DFT+DMFT calculation is around 4.42 $\mu_B$, which agrees well with the neutron diffraction measurement ($\sim$4.16 $\mu_B$).\cite{hill2008neutron} 
As demonstrated in Appendix Fig.~\ref{fig:band_temp}, at higher temperature of about 1450 K, the altermagnetic splitting can still be observed, especially for those bands that are close to the Fermi level, while the bands below get more strongly smeared. It should be noted that it has been known that the density-density form of Coulomb interaction would overestimate the magnetic transition temperature.\cite{belozerov2013magnetism} Therefore, the obtained local magnetic moment of Fe at $T=1450$ K is only slightly reduced to 4.22 $\mu_B$. 

\subsection{Altermagnetism and weak ferromagnetism}
\subsubsection{\label{sec:sym}Symmetry analysis}
Following the definition used by Dzyaloshinskii,\cite{dzyaloshinskii1957thermodynamic, borovik2006magnetic} $\alpha$-Fe$_2$O$_3$ belongs to the crystallographic space group $D_{3 d}^6=R \overline{3} c$. 
The representatives of the cosets $D_{3 d}^6 / \mathcal{T}$ then form the group $\widetilde{\boldsymbol{D}}_{3 d}^6$, which has the following elements:
\begin{equation}
	E, 2 C_3, 3 U_2, \tilde{I}, 3 \tilde{\sigma}_d, 2 \tilde{S}_6 \quad\left\{1, \pm 3_z, 3\left(2_{\perp}\right), \tilde{\overline{1}}, 3(c=\tilde{m}), \pm \tilde{\overline{3}}_z\right\}.
\end{equation}
Here, the sign $\tilde{}$ means that a translation along the $z$ axis through half the period of the crystal accompanies the corresponding symmetry operation. Following the definition in Ref.~\onlinecite{borovik2006magnetic} by marking the magnetic moments located at the points of the Fe ion positions as $\mu_\alpha$ with $\alpha=1,2,3,4$, the following linear combinations of $\mu_\alpha$ form the irreducible representations of $\widetilde{\boldsymbol{D}}_{3 d}^6$:
\begin{equation}
	\begin{aligned}
		& \mathbf{l}_1=\boldsymbol{\mu}_1+\boldsymbol{\mu}_2-\boldsymbol{\mu}_3-\boldsymbol{\mu}_4 \\
		& \mathbf{l}_2=\boldsymbol{\mu}_1-\boldsymbol{\mu}_2+\boldsymbol{\mu}_3-\boldsymbol{\mu}_4 \\
		& \mathbf{l}_3=\boldsymbol{\mu}_1-\boldsymbol{\mu}_2-\boldsymbol{\mu}_3+\boldsymbol{\mu}_4 \\
		& \mathbf{m}=\boldsymbol{\mu}_1+\boldsymbol{\mu}_2+\boldsymbol{\mu}_3+\boldsymbol{\mu}_4,
	\end{aligned}
\end{equation}
in which $\mathbf{l}_3$ represents the magnetic configuration as shown in Fig.~\ref{fig:band}(a). In the following analysis, we will then focus on the vectors $\mathbf{l}_3$ and $\mathbf{m}$. Table 1.5.3.2 of Ref.~\onlinecite{borovik2006magnetic} listed the sign variation of the components of magnetic vectors during transformations of the group $\widetilde{\boldsymbol{D}}_{3 d}^6$. Here, Table~\ref{tab:sym} was given for a recap with a slight modification on the symmetry operation element $2_x$ altered to $2_y$, following the definition of the basis vectors of $\alpha$-Fe$_2$O$_3$ in FPLO. 

\begin{table}[ht!]
	\renewcommand{\arraystretch}{1.25}
	\caption{\label{tab:sym} Sign variation of the components of $\mathbf{l}_3$ and $\mathbf{m}$ during transformations of the group $\widetilde{\boldsymbol{D}}_{3 d}^6$ in $\alpha$-Fe$_2$O$_3$}
\begin{tabular}{l|c|c|c|c|c|c|c|c|c|c}
	\hline
	\hline\multirow{3}{*}{} & \multicolumn{10}{c}{Elements of symmetry} \\
	\cline{2-11}
	Vector & E & $2 C_3$ & $U_2^1$ & $U_2^2$ & $U_2^3$ & $\tilde{I}$ &  $\tilde{\sigma}_d^1$ & $\tilde{\sigma}_d^2$ & $\tilde{\sigma}_d^3$ & $2 \tilde{S}_6$ \\
	components & 1 & $\pm 3_z$ & $2_y$ & $2_{\perp}^{(2)}$ & $2_{\perp}^{(3)}$ & $\tilde{\overline{1}}$ &  $c_y$ & $c_{\perp}^{(2)}$ & $c_{\perp}^{(3)}$ & $\pm \tilde{\overline{3}}$ \\
	\hline $l_{3 x}$ & + & 0 & + & 0 & 0 & + & + & 0 & 0 & 0 \\
	\hline $l_{3 y}$ & + & 0 & -- & 0 & 0 & + & -- & 0 & 0 & 0 \\
	\hline $l_{3 z}$ & + & + & + & + & + & + & + & + & + & + \\
	\hline $m_{x}$ & + & 0 & -- & 0 & 0 & + & -- & 0 & 0 & 0 \\
	\hline $m_{y}$ & + & 0 & + & 0 & 0 & + & + & 0 & 0 & 0 \\
	\hline $m_{z}$ & + & + & -- & -- & -- & + & -- & -- & -- & + \\
	\hline
\end{tabular}
\end{table}

In Table~\ref{tab:sym}, the + sign indicates that the element is included in the magnetic group, while for -- sign the element must be multiplied by time-inversion. The elements that yield 0 are not included in the magnetic group. Accordingly, when the magnetic moments are directed along $x$, $y$ and $z$ directions with vector $l_{3}$ (with corresponding magnetic structures denoted by $A_{3x}$, $A_{3y}$, and $A_{3z}$), the corresponding magnetic points groups are $2/m$, $2^\prime/m^\prime$, and $\overline{3} m$, respectively. For the ferromagnetic vector $m$, the magnetic points groups are $2^\prime/m^\prime$, $2/m$, and $\overline{3} m^\prime$, for the magnetic moments directed along $x$, $y$ and $z$, respectively (with corresponding magnetic structures denoted by $F_{x}$, $F_{y}$, and $F_{z}$). That is, the antiferromagnetic structure $A_{3x}$ ($A_{3y}$) and ferromagnetic structure $F_{y}$ ($F_{x}$) share the same magnetic point group. From a symmetry standpoint, the crystal structure permits the coexistence of antiferromagnetic and ferromagnetic components, thereby accounting for the weak ferromagnetism observed in $\alpha$-Fe$_2$O$_3$ above the Morin transition temperature, where the magnetic moments lie within the basal plane.

\begin{figure*}[ht!]
	\centering
	\includegraphics[width=1.0\linewidth]{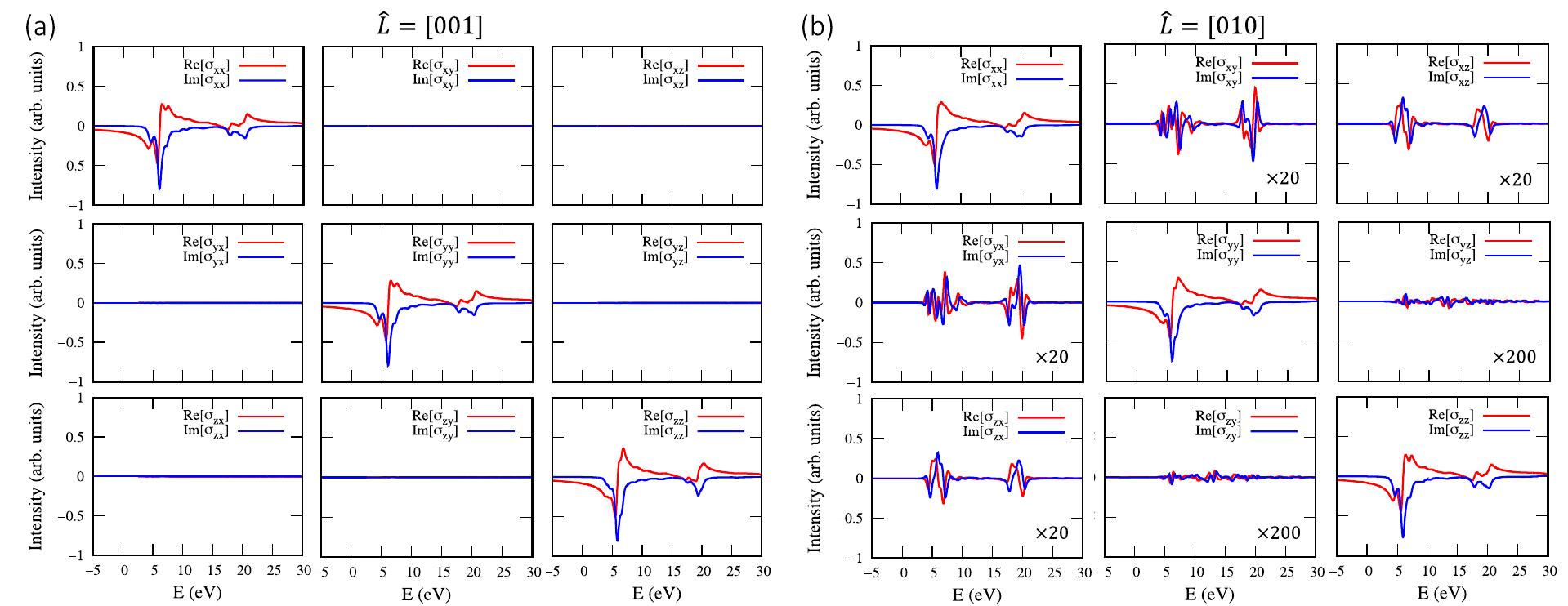}
	\caption{\label{fig:cond_t}X-ray absorption spectroscopy of Fe $\mathrm{L}_{2,3}$ edge in the form of a conductivity tensor for N\'eel vectors (a) $\hat{L}=[001]$ and (b) $\hat{L}=[010]$. The red and blue curves represent the real and imaginary parts, respectively.}
\end{figure*}

\begin{figure}[h!]
	\centering
	\includegraphics[width=0.72\columnwidth]{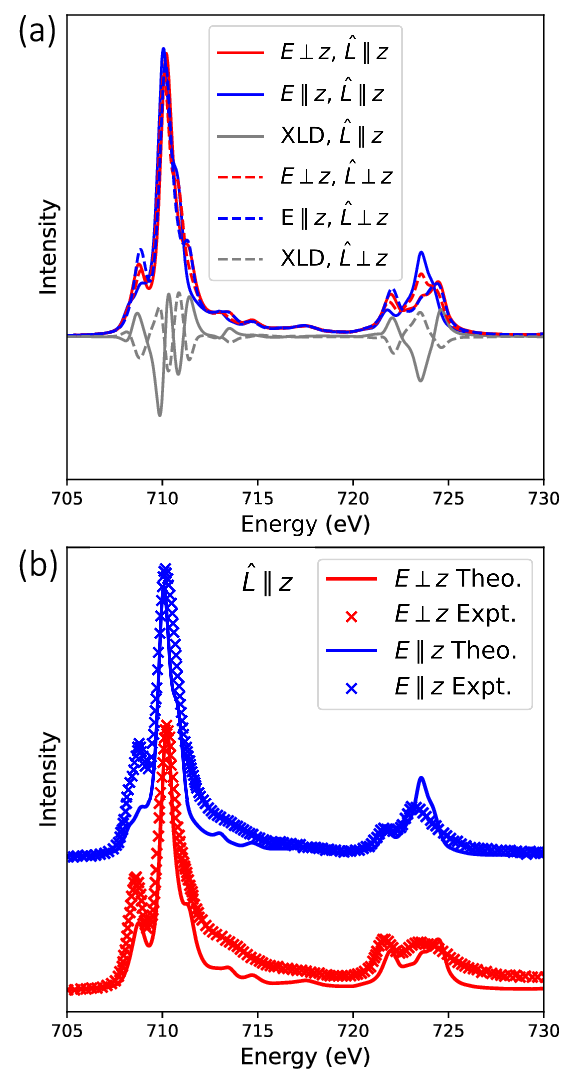}
	\caption{\label{fig:xld}(a) Fe 2p XAS of $\alpha$-Fe$_2$O$_3$ for N\'eel vector $\hat{L}$ along $z$ with x-ray polarization perpendicular and parallel to $\hat{L}$. (b) X-ray linearized dichroism (XLD) for $\hat{L}$ perpendicular and parallel to $z$.}
\end{figure}

Inspired by the XMCD distinction between altermagnetism and weak ferromagnetism reported in $\alpha$-MnTe\cite{hariki2024x}, we apply a similar analysis to $\alpha$-Fe$_2$O$_3$, additionally focusing on the spectroscopic variation below and above the Morin transition temperature. 
Below the Morin transition temperature, the N\'eel vector is along $z$ direction ($\hat{L}=[001]$) with magnetic point group $\overline{3} m$. 
Following the symmetry-adapted form of tensor properties for $\overline{3} m$, the table of tensor components has the following form:
\begin{equation}
	\label{eq:z_tensor}
	\left[\begin{array}{lll}
		\sigma_{x x} & 0 & 0 \\
		0 & \sigma_{x x} & 0 \\
		0 & 0 & \sigma_{z z}
	\end{array}\right].
\end{equation}
The mirror planes containing the $z$-axis forbid any off-diagonal (Hall-type) components in the conductivity tensor. As demonstrated in Fig.~\ref{fig:cond_t}(a), the off-diagonal terms of XAS conductivity tensor are zero, and the diagonal components exhibit that $\sigma_{x x}=\sigma_{y y}\neq\sigma_{z z}$. 
We emphasize that the spectra reported in this work are normalized to the contributions from the four Fe atoms in the primitive cell, which is allowed by the local nature of core-level excitations.\cite{winder2020x}
The symmetry analysis for the magnetic point group $\overline{3} m$ indicates that there exists no XMCD signal when the N\'eel vector is along [001] direction. However, the x-ray linearized dichroism (XLD) persists because of the intrinsic uniaxial anisotropy as indicated by Eq.~\ref{eq:z_tensor} (see Fig.~\ref{fig:xld}). Additionally, Fig.~\ref{fig:xld}(b) shows the calculated XASs when $\hat{L}=[001]$ with x-ray polarization perpendicular and parallel to [001] direction, which are in good agreement with experimental measurements.\cite{kuiper1993x}

Above the Morin transition temperature, the N\'eel vector switches its direction towards [010], with the magnetic point group belonging to $2^{\prime}/m^{\prime}$. The corresponding symmetry-adapted tensor has the form
\begin{equation}
	\label{eq:z_tensor2}
	\left[\begin{array}{lll}
		\sigma_{x x} & \sigma_{xy} & \sigma_{xz} \\
		-\sigma_{xy} & \sigma_{y y} &  \sigma_{yz}  \\
		 \sigma_{xz} & -\sigma_{yz} & \sigma_{z z}
	\end{array}\right].
\end{equation}
The calculated tensor form of XAS for $\hat{L}=[010]$ as shown in Fig.~\ref{fig:cond_t}(b) well coincides with the form described by Eq.~\ref{eq:z_tensor2}. There are six independent components $\sigma_{x x} $, $\sigma_{yy} $, $\sigma_{zz} $, $\sigma_{xy}$, $\sigma_{xz}$ and $\sigma_{yz}$. The conductivity tensor contains the antisymmetric part, i.e., $\sigma_{yx}=-\sigma_{xy}$, and $\sigma_{zy}=-\sigma_{yz}$. This indicates that for $\hat{L}=[010]$, when the x-ray propagation vector $\hat{k}$ is along $z$ or $x$ direction, the XMCD response should be expected. 
In addition, the XLD signal is also present, and the spin reorientation across the Morin transition causes a sign reversal of the dichroic signal (see Fig.~\ref{fig:xld}(a)), as also observed experimentally in the work of Refs.~\onlinecite{kuiper1993x, jani2021reversible}. 

\subsubsection{\label{sec:xmcd}XMCD}
\begin{figure*}[ht!]
	\centering
	\includegraphics[width=1\linewidth]{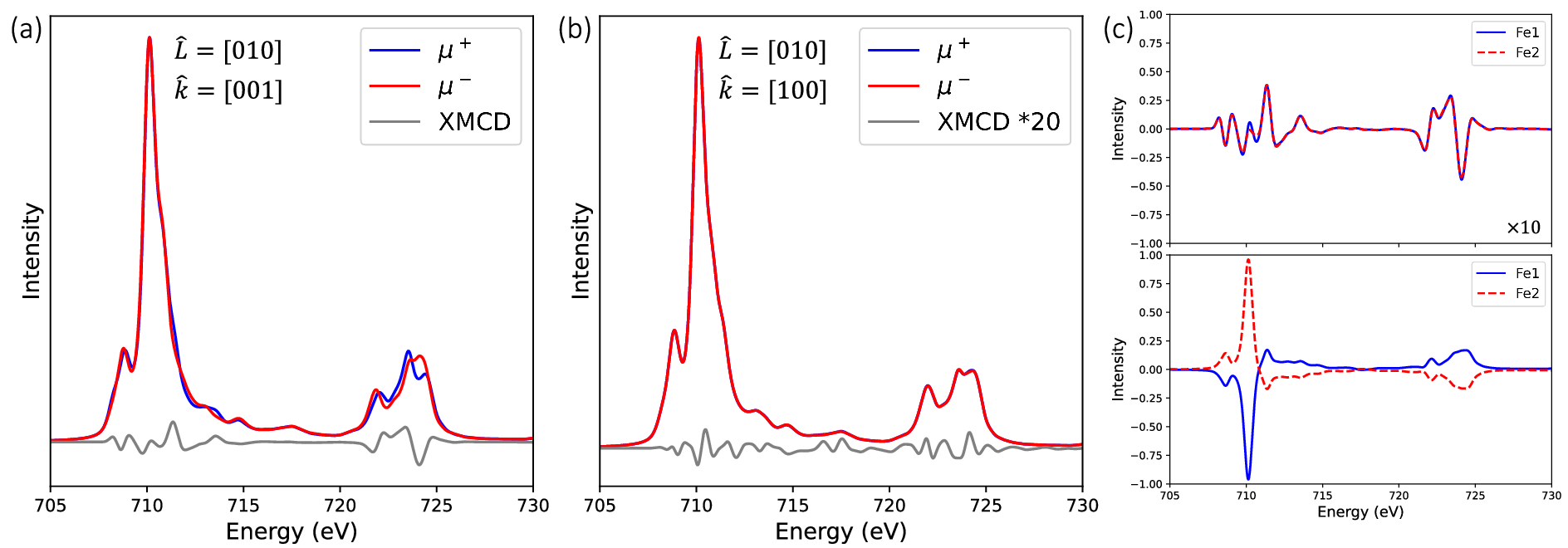}
	\caption{\label{fig:xmcd}Fe 2p XMCD spectra for $\hat{L}=[010]$ when the x-ray propagation vector $\hat{k}$ is along (a)[001] and (b)[100] directions. (c)Upper panel: the separate contributions from Fe1 and Fe2 to XMCD with $\hat{L}=[010]$ and $\hat{k}=[001]$. Lower panel: calculated XMCD of Fe1 and Fe2 at applied field of 2 T for $\hat{k}=[010]$, which is parallel to the N\'eel vector $\hat{L}=[010]$.}
\end{figure*}
As indicated by the symmetry analysis in Sec.~\ref{sec:sym} and as demonstrated in Fig.~\ref{fig:xmcd}(a) and (b), with $\hat{L}=[010]$, XMCD responses are present for the x-ray propagation vectors $\hat{k}$ along [001] and [100] directions. However, the XMCD response is rather weak for $\hat{k}=[100]$, which would in practice hardly detectable. 
Note that in these two cases, the x-ray propagation vector is perpendicular to the direction of spin moment, which is not a common application for XMCD measurements in ferromagnets. 
Subsequently, using the setup $\hat{L}=[010]$, and $\hat{k}=[001]$, the atom-resolved XMCD spectra are shown in Fig.~\ref{fig:xmcd}(c). For clarity we only show the contributions from Fe1 and Fe2 atoms, because as an inversion pair of Fe1 and Fe2, Fe4 and Fe3 possess the same crystal field as Fe1 and Fe2, respectively. Therefore, the XMCD spectra from Fe1 and Fe4 (Fe2 and Fe3) are identical. 

It can be seen from the upper panel of Fig.~\ref{fig:xmcd}(c) that for $\hat{k}=[001]$ with $\hat{L}=[010]$, Fe1 and Fe2 contribute almost equally to the XMCD, in which the slight deviation should be due to numerical noise. As shown in Eq.~\ref{eq:z_tensor2}, the bulk tensor $\sigma_{xy}$ demonstrates antisymmetry. We further decompose $\sigma_{xy}$ into site-resolved contributions. Interestingly, it is found that the site-resolved $\sigma_{xy}^i$ (with $i$ denoting the site index) contains both symmetry ($S_{xy}^i=\frac{1}{2}(\sigma_{xy}^i+\sigma_{yx}^i)$ ) and antisymmetry ($A_{xy}^i=\frac{1}{2}(\sigma_{xy}^i-\sigma_{yx}^i)$) contributions. That is, although the bulk symmetry suppresses the symmetry part of $\sigma_{xy}$, locally it is allowed and finite. Since Fe2 is connected to Fe1 by $C_{2y}t_{1/2}$ symmetry operation, $S_{xy}^1 = -S_{xy}^2$ and $A_{xy}^1 = A_{xy}^2$. This statement is also valid for $\sigma_{yz}$. In terms of $\sigma_{xz}$ that shows symmetric behaviour ($S_{xz}^1 = S_{xz}^2$), its site-resolved antisymmetric component ($A_{xz}^i$) has a quite prominent magnitude although $A_{xz}^1$ and $A_{xz}^2$ cancel out as an integrity (see Appendix Fig.~\ref{fig:anti}). Consequently, a small symmetry breaking perturbation can instantly release those locally present but globally cancelled contributions.

In order to differentiate the contributions from altermagnetism and weak ferromagnetism to XMCD in $\alpha$-Fe$_2$O$_3$, we also plotted out the site-resolved XMCD response for $\hat{L}=[010]$ and $\hat{k}=[010]$ in the lower panel of Fig.~\ref{fig:xmcd}(c). Here, an external magnetic field of around 2 T was applied. A large difference on the XMCD spectral line shape can be observed for $\hat{k}=[010]$ and $\hat{k}=[001]$. Additionally, when $\hat{k}\parallel\hat{L}$, the XMCD responses from Fe1 and Fe2 have opposite sign, while they have the same sign for $\hat{k}\perp\hat{L}$. This phenomenon has also been reported in altermagnetic $\alpha$-MnTe\cite{hariki2024x}. Another difference between the two setups $\hat{k}\parallel\hat{L}$ and $\hat{k}\perp\hat{L}$ is that for $\hat{k}\perp\hat{L}$ the XMCD response is strongly dependent on the direction of $\hat{L}$. For instance, as demonstrated in Fig.~\ref{fig:cond_t}(a) and Fig.~\ref{fig:Lx_cond} (see Appendix C), for $\hat{L}=[001]$ and $\hat{L}=[100]$ that correspond to magnetic point group $\overline{3}m$ and $2/m$ respectively, the XMCD signals are forbidden by symmetry.

 \subsubsection{Discussion}
 
  \begin{figure}[h!]
 	\centering
 	\includegraphics[width=1\columnwidth]{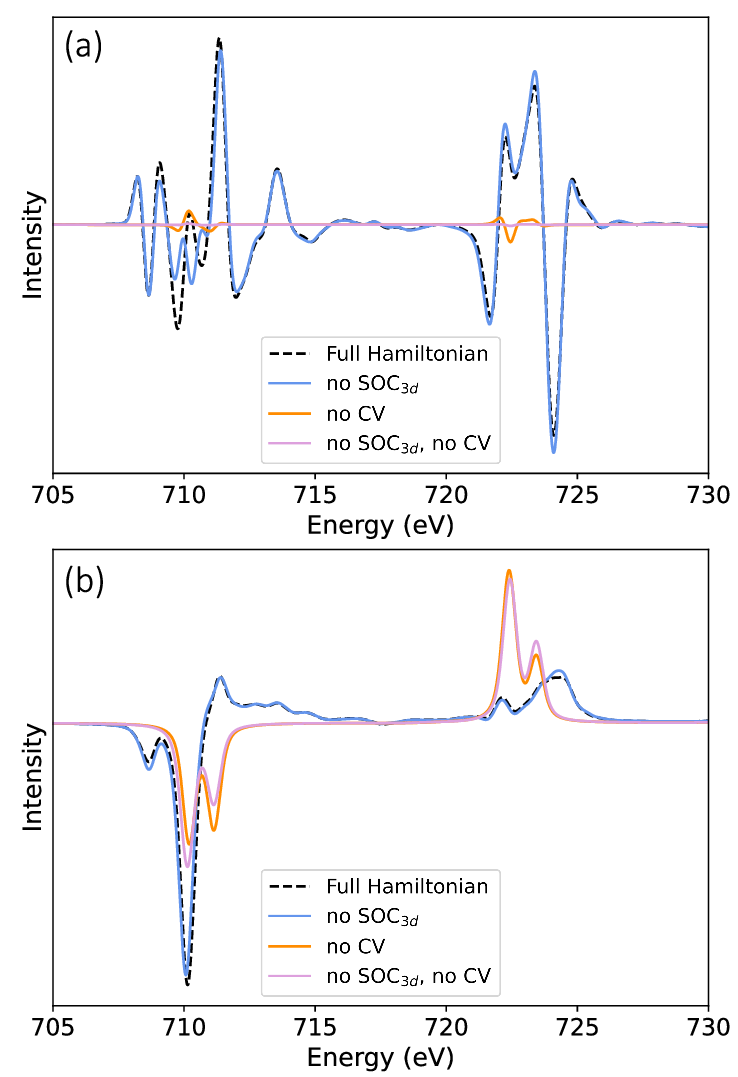}
 	\caption{\label{fig:xmcd_ana}The single-site Fe 2p XMCD spectra for $\hat{L}=[010]$ with the x-ray propagation vector (a) $\hat{k}=[001]$ and (b) $\hat{k}=[010]$. The contributions originated from the SOC effect of 3d valence states, the core-valence (CV) beyond monopole interaction are explicitly illustrated.}
 \end{figure}
 
 Furthermore, we performed a more detailed analysis on the obtained single-site XMCD spectrum for $\hat{L}\perp\hat{k}$ to inspect the origin of XMCD response. It has been shown in the work of Hariki et al. that for altermagnetic $\alpha$-MnTe the valence SOC has a negligible effect on its XMCD, whereas XMCD response disappears when a combination of core SOC and multipole core-valence exchange interactions (CV) are absent.\cite{hariki2024x} Nevertheless, this observation is system-dependent based on the specific symmetry existed in the crystal.\cite{kunevs2025neel} For instance, in altermagnetic MnF$_2$, the XMCD is still present even if the core SOC and CV terms are excluded.\cite{hariki2024determination} In the present work, we provide the symmetry analysis of the Hamiltonian in Eq.~\ref{eq:ctm} for $\alpha$-Fe$_2$O$_3$ following the argument in Ref.~\onlinecite{hariki2024x}. By excluding the terms defining core SOC and multipole CV exchange interactions 
 \begin{equation}
 	H_{\mathrm{at}}^1=\sum_{\substack{i, j, k, l \\ \sigma, \sigma^{\prime}}} W_{i j k l}^{\sigma \sigma^{\prime}} \boldsymbol{d}_{i \sigma}^{\dagger} \boldsymbol{p}_{j \sigma^{\prime}}^{\dagger} \boldsymbol{d}_{k \sigma^{\prime}} \boldsymbol{p}_{l \sigma}+\sum_{\substack{m, m^{\prime} \\ \sigma, \sigma^{\prime}}} h_{m \sigma, m^{\prime} \sigma^{\prime}}^{(3d)} \boldsymbol{d}_{m \sigma}^{\dagger} \boldsymbol{d}_{m^{\prime} \sigma^{\prime}}
 \end{equation}
 from Eq.~\ref{eq:ctm}, the remaining terms constitute the following Hamiltonian:
\begin{equation}
\label{eq:h0}
 \begin{aligned}
 \!\!\!\!	H_{\mathrm{at}}^0 = \sum_{\substack{i, j, k, l \\ \sigma, \sigma^{\prime}}} U_{i j k l}^{\sigma \sigma^{\prime}} \boldsymbol{d}_{i \sigma}^{\dagger} \boldsymbol{d}_{j \sigma^{\prime}}^{\dagger} \boldsymbol{d}_{k \sigma^{\prime}} \boldsymbol{d}_{l \sigma} + U_{p d} \boldsymbol{n}_p \boldsymbol{n}_d + \\
\!\!\!\! 	 \sum_{m, m^\prime, \sigma, \sigma^{\prime}} h_{m \sigma, m^{\prime} \sigma^{\prime}}^{(2 \mathrm{p})} \boldsymbol{p}_{m \sigma}^{\dagger} \boldsymbol{p}_{m^{\prime}\sigma^{\prime}} +  \sum_{m, m^{\prime}, \sigma} h_{m m^{\prime}}^{(\mathrm{CF})} \boldsymbol{d}_{m \sigma}^{\dagger} \boldsymbol{d}_{m^{\prime} \sigma} + \\
\!\!\!\!  \sum_{L} \sum_{m, m^{\prime}, \sigma} h_{L, m m^{\prime}}^{(\mathrm{CF})} \boldsymbol{L}_{m \sigma}^{\dagger} \boldsymbol{L}_{m^{\prime} \sigma} + 
 	 (\sum_{m,m^\prime,\sigma} V_{m m^\prime}^{(dL)}\boldsymbol{d}_{m \sigma}^{\dagger} \boldsymbol{L}_{m^{\prime}\sigma} + H.c.) \\
\!\!\!\!  +(\sum_{L}^{n_L-1} \sum_{m, m^{\prime}, \sigma}V_{m m^\prime}^{L,L+1}\boldsymbol{L}_{m \sigma}^{\dagger} \boldsymbol{L}_{m^{\prime}\sigma} + H.c.) \\
 \!\!\!\! + \sum_{m, m^{\prime}, \sigma, \sigma^{\prime}} B_{m \sigma, m^{\prime} \sigma^{\prime}} \boldsymbol{d}_{m \sigma}^{\dagger} \boldsymbol{d}_{m^{\prime} \sigma^{\prime}}.
 \end{aligned}
\end{equation}
We would like to mention that as compared to Ref.~\onlinecite{hariki2024x}, the definition of Weiss field term $\sum\limits_{m, m^{\prime}, \sigma, \sigma^{\prime}} B_{m \sigma, m^{\prime} \sigma^{\prime}} \boldsymbol{d}_{m \sigma}^{\dagger} \boldsymbol{d}_{m^{\prime} \sigma^{\prime}}$ in Eq.~\ref{eq:h0} is slightly different, which includes also the orbital moment besides the spin moment. 
In the following symmetry analysis, the consequences brought by the presence of the orbital momentum will be discussed. 

First, we show in Fig.~\ref{fig:xmcd_ana} the single-site XMCD of Fe1 in $\alpha$-Fe$_2$O$_3$ with $\hat{k}\perp\hat{L}$ ($\hat{k}=[001]$, $\hat{L}=[010]$). By manually eliminating the valence SOC term in $H_{\mathrm{CTM}}$, the XMCD response is only slightly modified. In contrast, if the core-valence Coulomb interaction terms beyond monopole were removed, the remaining XMCD signal becomes very tiny. Finally, when both valence SOC and CV terms are withdrew, the XMCD response almost disappears. 

This observation can be properly rationalized from a perspective of symmetry. 
As expressed in Eq.~\ref{eq:xmcd_g} , the spectrum object can be equivalently written as 
\begin{equation}
	F_{ \pm}(\omega)=\sum_f|\langle f| T_{ \pm}| i \rangle |^ 2 \delta(\omega-(E_f-E_i))
\end{equation}
with the transition operator for obtaining XMCD defined by
\begin{equation}
	\label{eq:dipo}
	T_{ \pm} \equiv \sum_\sigma T_{ \pm}^\sigma=\sum_{m, \sigma} \Gamma_{ \pm m} \boldsymbol{d}_{m \pm 1 \sigma}^{\dagger} \boldsymbol{p}_{m \sigma}+\text { H.c.},
\end{equation}
describing absorption of circularly polarized light propagating along the $z$ axis ($\hat{k}=[001]$). Here, $d_{m \pm 1 \sigma}^{\dagger} $ ($p_{m \sigma}$) create an electron (a hole) in valence (core) orbital with angular momentum projection $m \pm 1 $ ($m$) and spin projection $\sigma$. The square of the matrix element in Eq.~\ref{eq:dipo} is then
\begin{equation}
	\label{eq:T_2}
	\left|\left\langle T_{ \pm}\right\rangle\right|^2=\sum_\sigma\left|\left\langle T_{ \pm}^\sigma\right\rangle\right|^2+\sum_\sigma\left\langle T_{ \pm}^\sigma\right\rangle \overline{\left\langle T_{ \pm}^{-\sigma}\right\rangle}.
\end{equation}
It has been shown for $\alpha$-MnTe that the symmetry operation $C_{3}$ commutes with the Hamiltonian that omits the valence SOC and CV terms. Under such circumstance, the second term in Eq.~\ref{eq:T_2} drops out considering the $C_{3}$ transformation acting only on the valence orbital indices and both core spin and orbital indices.\cite{hariki2024x} Together with the vanishing crossed terms in Eq.~\ref{eq:T_2}, the antiunitary symmetry $\mathcal{T}^{\prime} \equiv \mathcal{T} \mathcal{C}_2^{S, 3 d}$ (time-reversal combined with rotation of the valence spin around an axis perpendicular to the ordered moments) leads to $F_{+}(\omega)=F_{-}(\omega)$, hence zero XMCD in $\alpha$-MnTe.\cite{hariki2024x}

Similar to $\alpha$-MnTe, the $\alpha$-Fe$_2$O$_3$ crystal demonstrates $C_{3z}$ symmetry (see Table~\ref{tab:sym}). Accordingly, in the angular momentum basis, the matrix elements of crystal-field Hamiltonian $h_{m m^{\prime}}^{(\mathrm{CF})}$ which do not fulfil the relation that $m-m^\prime=3k$, should be zero. The $h_{m m^{\prime}}^{(\mathrm{CF})}$ extracted from FPLO calculation indeed verifies this symmetry relation (See Appendix Eq.~\ref{eq:h_cf}). Similarly, the crystal-field Hamiltonians for the ligands of the same size as $3d$ orbital $h_{L, m m^{\prime}}^{(\mathrm{CF})}$, as well as their hybridization terms, all follow the same symmetry, since in the applied MLFT, the Hamiltonian corresponding to the selected cluster is rotated using an unitary transformation so that the consequent representation defines a one-dimensional chain of interactions. In other words, the site (3$d$ orbital of Fe) under consideration interacts only with one ligand orbital directly, which in turn interacts with another ligand.\cite{lu2014efficient} In addition, there is a term that is not explicitly visible in Eq.~\ref{eq:h0} but considered in DFT + MLFT calculations, i.e., the correction for screening effect which has been applied by deducting the on-site energies of Fe $2p$, $3d$ and ligand orbitals obtained following the convention used in the work of Bocquet et al.\cite{bocquet1992electronic} 

We then consider the $C_{3z}$ transformation acting only on the valence orbital indices, but for core on both orbital and spin indices, as performed in Ref.~\onlinecite{hariki2024x}. Due to the presence of orbital moment in Eq.~\ref{eq:h0} for $\hat{L}=[010]$, the $C_{3z}$ symmetry is actually broken for a single magnetic domain. However, since the orbital contribution is rather small in Fe, the XMCD response resulted from $H_{\mathrm{at}}^0$ is close to vanishing, as indicated by the pink line in Fig.~\ref{fig:xmcd_ana}. If the orbital part in the Weiss field term were further omitted, the commute relation between $C_{3z}$ transformation and $H_{\mathrm{at}}^0$ will be mostly recovered. It is not fully recovered because with the employed DFT + MLFT method in this work, the rotated cluster Hamiltonian with the form of one-dimensional chain interactions does not only give ligand orbitals possessing the same size and symmetry as $3d$ orbital, but also ligand orbitals of the size of $s$ orbital. 

Another point worth discussing is whether the XMCD response can be detected in actual experiments. In the case of $\alpha$-MnTe,\cite{hariki2024x} the shapes of the calculated and measured XMCD spectra agree rather well, but the measured XMCD intensity is about 10 times smaller than the calculated result. Such magnitude difference was attributed to the presence of magnetic domains with N\'eel vectors pointing towards the six easy axis directions. On the one hand, this argument is still valid for $\alpha$-Fe$_2$O$_3$. On the other hand, the XMCD of altermagnetic $\alpha$-Fe$_2$O$_3$ is only allowed above Morin transition temperature. Therefore, we additionally carried out XMCD simulations at 250 K, incorporating the expectation values of two thermally populated excited states through Boltzmann weighting. At $T=250$ K, the XMCD magnitude is further reduced. Consequently, detecting the XMCD response experimentally at temperatures above the Morin transition temperature would be highly challenging, especially in the presence of unavoidable instrumental noise.

\section{\label{sec:con}Conclusions}
In conclusion, despite the different band features demonstrated by DFT, DFT+U and DFT+DMFT approaches, the characteristic spin splitting in altermagnetic $\alpha$-Fe$_2$O$_3$ persists. Additionally, the spin splitting near the Fermi level is still very apparent and only gets slightly smeared at high temperatures, as indicated by the finite-temperature DFT+DMFT calculations. Moreover, the spin splitting is found to be insensitive to the spin orientation with SOC included. 
Below the Morin transition temperature, the N\'eel vector is along [001] direction and the crystal belongs to the $\overline{3} m$ magnetic point group. For this configuration, only diagonal terms of the conductivity tensors are allowed: $\sigma_{x x}=\sigma_{y y}\neq\sigma_{z z}$. In this regard, XLD is present and the calculated XLD agrees well with the experimental measurements. 
On the other hand, the XMCD response is allowed only above Morin transition temperature when the N\'eel vector is along [010] direction, which corresponds to the magnetic point group $2^\prime/m^\prime$. Interestingly, we find that the site-resolved contributions to the off-diagonal terms in the conductivity tensor for $\hat{L}=[010]$ configuration involves both symmetric and antisymmetric components, although the specific tensor form is restricted by the magnetic point group as a whole.
This observation indicates that a small symmetry breaking perturbation is able to instantly release those locally present but globally cancelled terms, which is an intriguing property that can be exploited via, e.g., doping. 

We would also like to emphasize that the $ab-initio$ DFT+MLFT approach adopted in this work is an effective and reliable tool to simulate the XMCD spectroscopy. Nevertheless, when performing symmetry analysis, one should pay extra attention to the hybridization terms between the valence orbitals of transition metal (TM) and its ligands. In specific, the hybridization terms within the considered TM-centred clusters might include ligands that do not possess the same symmetry as the TM site after unitary transformation. 
Through symmetry analysis, we demonstrate that the valence spin–orbit coupling plays a negligible role in the XMCD response of $\alpha$-Fe$_2$O$_3$, establishing XMCD as a promising probe of altermagnetism that does not rely on relativistic effects and therefore avoids ambiguities with weak ferromagnetism. Furthermore, the distinct XMCD spectral line shapes for light propagation vectors parallel and perpendicular to the N\'eel vector provide a clear means to disentangle altermagnetic contributions from weak ferromagnetic ones. While inclusion of temperature effects in the XMCD simulations is expected to reduce the overall signal intensity, our results nevertheless offer clear guidance for experimental conditions under which the altermagnetism-induced XMCD response can be identified.

\begin{acknowledgments}
The authors gratefully acknowledge the computing time provided to them at the NHR Center NHR4CES at RWTH Aachen University (project number p0024007). This is funded by the German Federal Ministry of Research, Technology, and Space (BMFTR), and the state governments participating on the basis of the resolutions of the Gemeinsame Wissenschaftskonferenz (GWK) for national high performance computing at universities.  
\end{acknowledgments}

\section*{Appendix A - Computational Details}
\begin{figure}[h!]
		\centering
		\includegraphics[width=1.0\columnwidth]{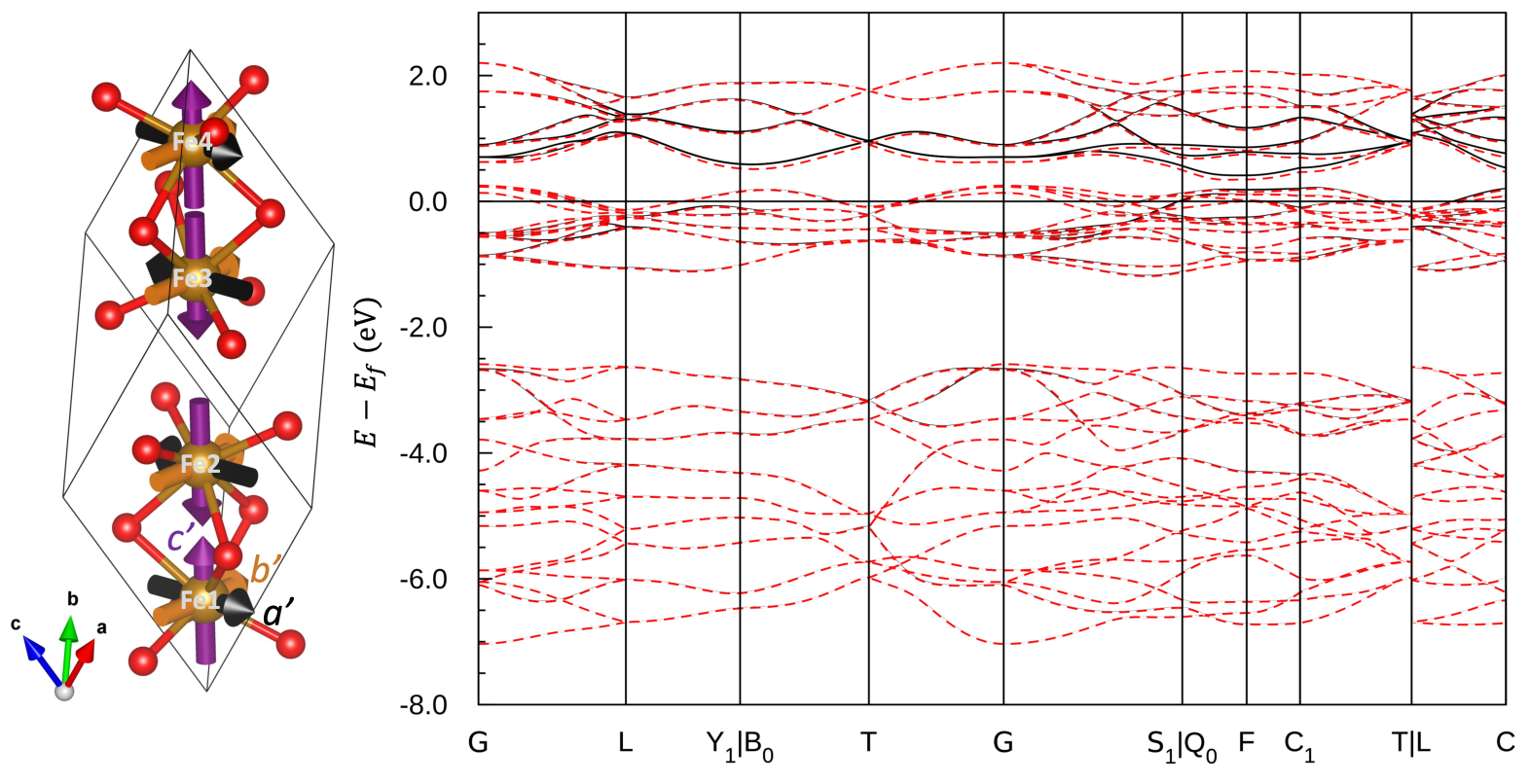}
		\caption{\label{fig:wann} Band structures of $\alpha$-Fe$_2$O$_3$ calculated using FPLO by DFT (black solid line) and Wannier functions (red dash line). Here, $a^{\prime}$, $b^{\prime}$ and $c^{\prime}$ define the local coordination axes (in Cartesian coordinates) of Fe atoms.}
	\end{figure}

As shown in Fig.~\ref{fig:wann}, the wannierization process in FPLO was performed with an energy window of [-8, 3] eV, in which the Fe 3$d$ and O 2$p$ orbitals are included for constructing the Wannier functions. The obtained band structure from Wannier functions agree well with that is directly calculated by DFT, thus validating the accuracy of using Wannier functions to further construct the tight-binding model.  In addition, the specific values of atomic Slater integral which are derived from the radial functions generated by FPLO are listed in Table.~\ref{tab:slater}.

\begin{table}[h!]
	\centering
	\caption{\label{tab:slater} The atomic Slater integral values derived from DFT radial functions.}
	\begin{tabular}{c|lllll}
		\hline\hline
		& $F_2^{dd}$ & $F_4^{dd}$ & $F_2^{pd}$ & $G_1^{pd}$ & $G_3^{pd}$ \\
		\hline
		Values $(\mathrm{eV})$ & 10.42 & 6.44 & 5.96 & 4.35 & 2.47 \\
		\hline\hline
	\end{tabular}
\end{table}

\section*{Appendix B - SOC and Temperature Effects}
The band structures of $\alpha$-Fe$_2$O$_3$  with N\'{e}el vectors along [001] and [010] directions are calculated using VASP in the regime of DFT+U+SOC. By projecting the spin angular momentum to the corresponding quantization axis, we find that the characteristic spin splitting in altermagnetic $\alpha$-Fe$_2$O$_3$ persists with the inclusion of SOC, or with different spin orientations in the presence of SOC. 

\begin{figure}[h!]
	\centering
	\includegraphics[width=1.0\columnwidth]{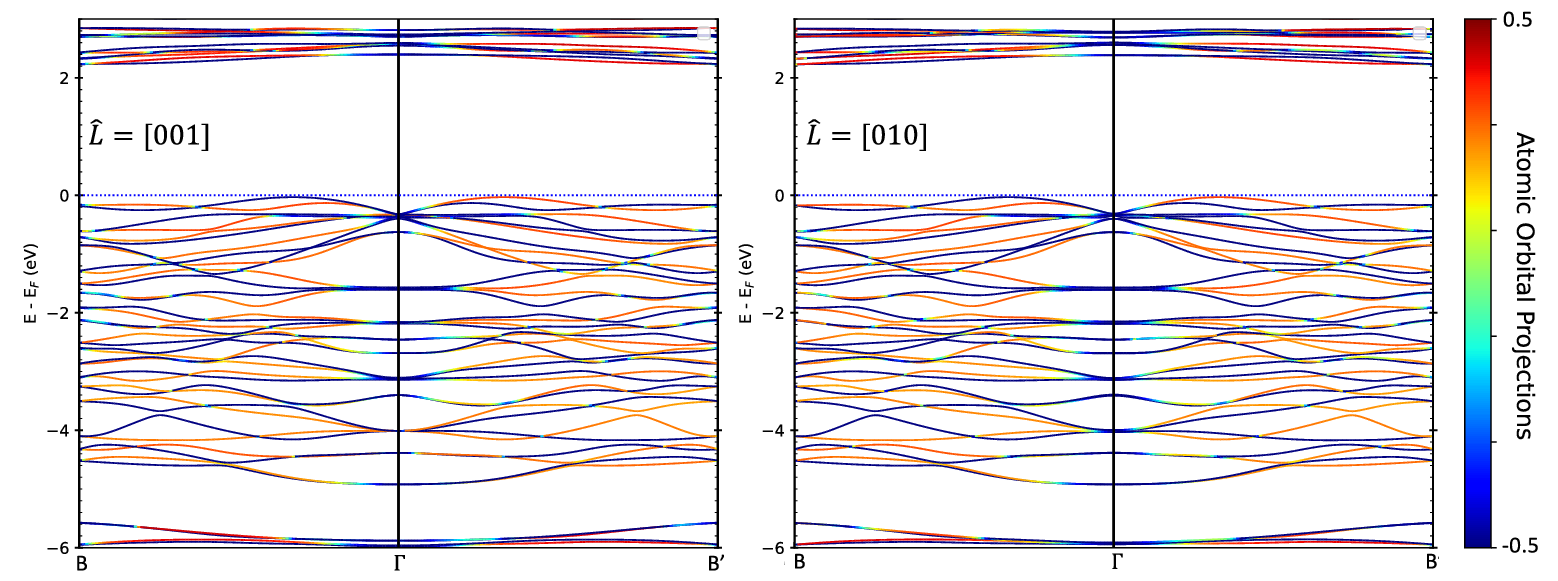}
	\caption{\label{fig:band_soc} Band structures of $\alpha$-Fe$_2$O$_3$ calculated using DFT+U+SOC for N\'{e}el vectors along [001] (left panel) and [010] (right panel) directions, respectively. The colour code is the projected spin momentum along the directions of the N\'{e}el vectors.}
\end{figure}

\begin{figure}[h!]
	\centering
	\includegraphics[width=1.0\columnwidth]{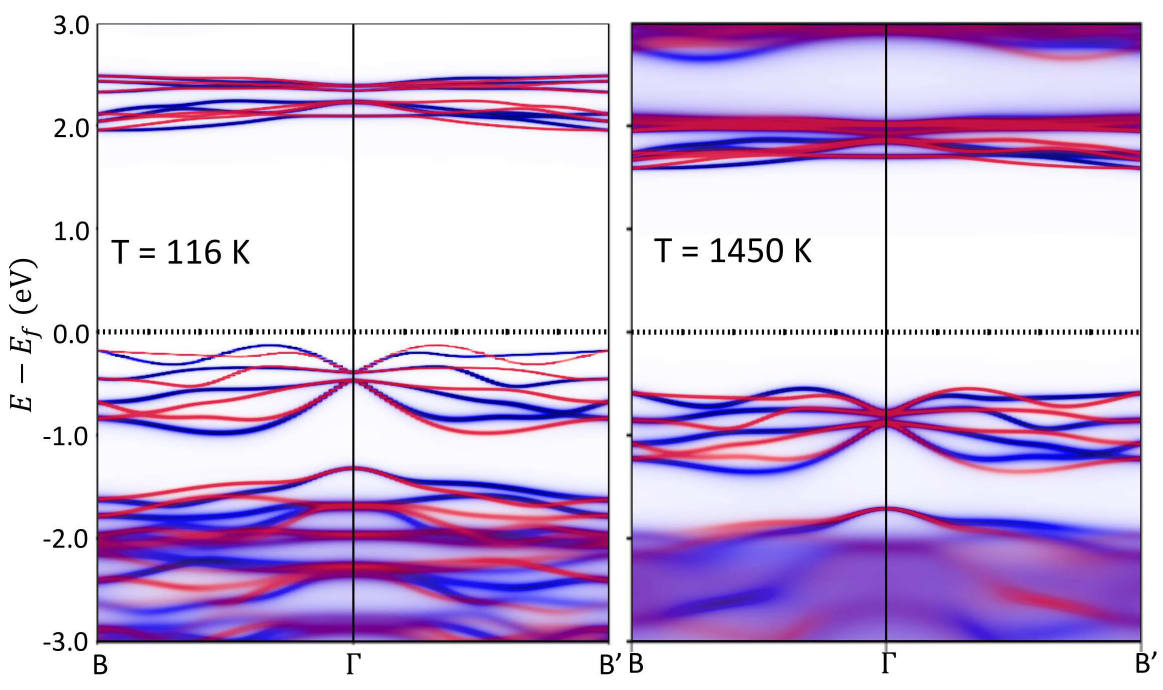}
	\caption{\label{fig:band_temp} Band structures of $\alpha$-Fe$_2$O$_3$ calculated using DFT+DMFT at $T=116$ K (left panel) and $T=1450$ K (right panel).}
\end{figure}

Furthermore, the band structure obtained using DFT+DMFT method at higher temperature $T=1450$ K is shown in Fig.~\ref{fig:band_temp} and has been discussed in Sec.~\ref{sec:band}

\begin{figure}[ht!]
	\centering
	\includegraphics[width=0.85\columnwidth]{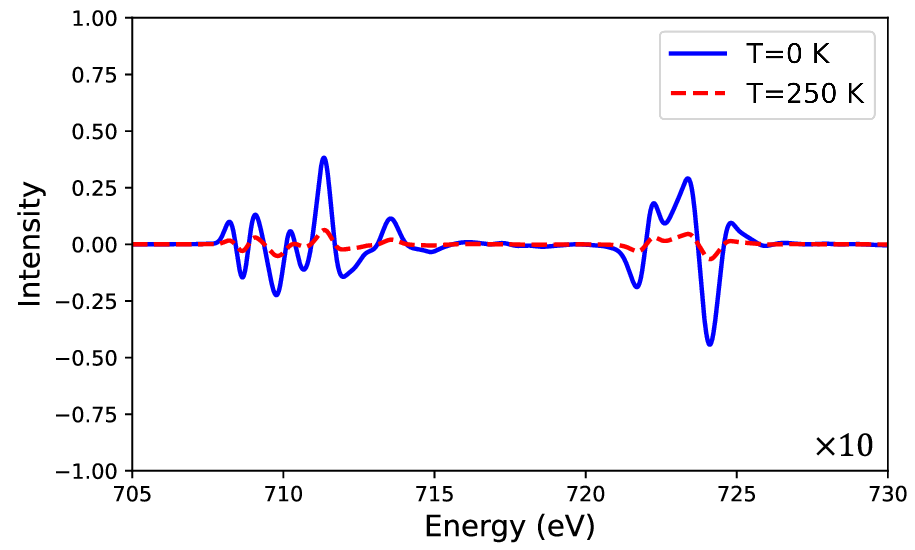}
	\caption{\label{fig:xmcd_temp} The XMCD spectra with $\hat{L}=[010]$ and $\hat{k}=[001]$ at $T=0$ K and $T=250$ K.}
\end{figure}

\section*{Appendix C - Additional Symmetry Relations}
\begin{figure}[h!]
	\centering
	\includegraphics[width=0.85\columnwidth]{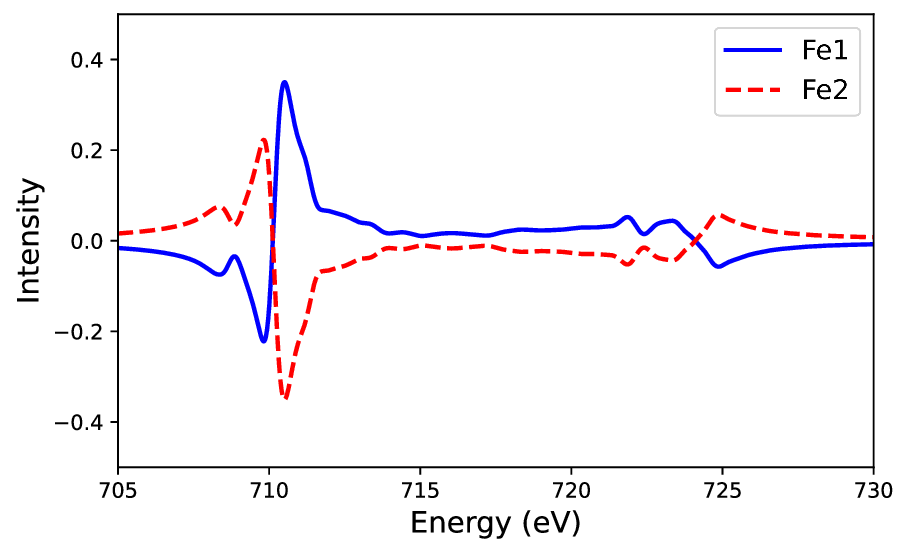}
	\caption{\label{fig:anti}The site-resolved antisymmetric part $A_{xz}^i$ ($i=1,2$ for atomic sites Fe1 and Fe2) for $\hat{k}=[001]$ with $\hat{L}=[010]$.}
\end{figure}

The demonstration of the site-resolved antisymmetric part $A_{xz}^i$ ($i=1,2$ for atomic sites Fe1 and Fe2) in Fig.~\ref{fig:anti} is to support the argument in Sec.~\ref{sec:xmcd} that although the bulk symmetry forbids the manifestation of the antisymmetric contribution ($\sigma_{xz}=\sigma_{zx}$), intrinsically the site-resolved contribution is not zero and can manifest itself if a symmetry breaking perturbation were introduced. 

\begin{figure}[h!]
	\centering
	\includegraphics[width=1.0\columnwidth]{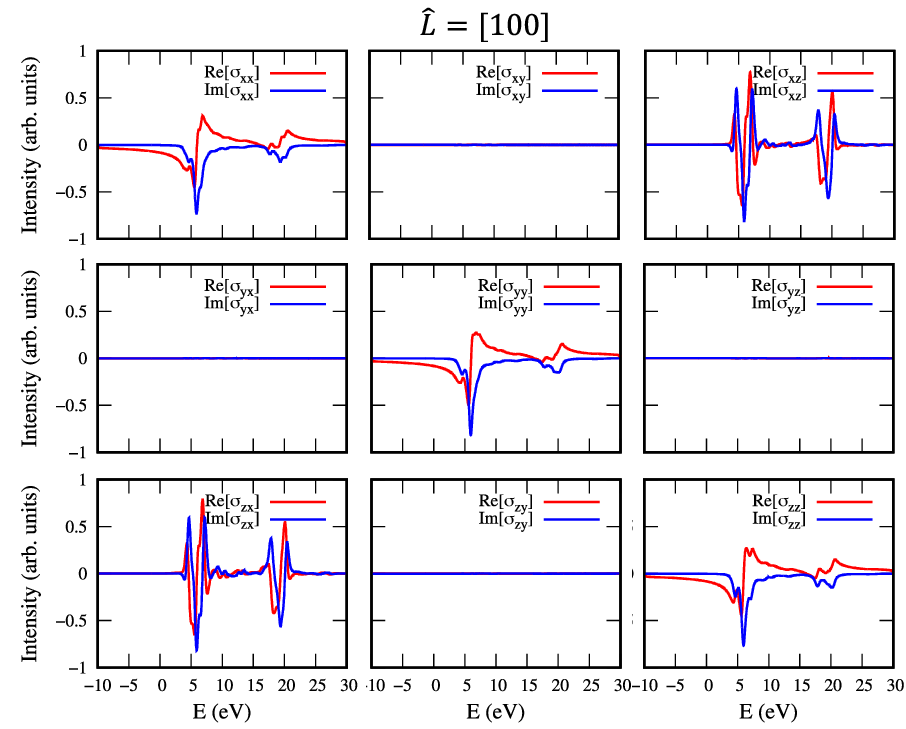}
	\caption{\label{fig:Lx_cond}X-ray absorption spectroscopy of Fe $\mathrm{L}_{2,3}$ edge in the form of conductivity tensor for N\'eel vector $\hat{L}=[100]$.}
\end{figure}

The specific form of $h_{mm^\prime}^{{CF}}$ can be extracted from the Wannier functions constructed in FPLO. However, it is originally expressed in the real basis and one needs to apply a basis transformation from real to the angular momentum basis.
\begin{equation}
		\label{eq:h_cf}
	\scriptsize
	\begin{aligned}
h_{mm^\prime}^{{CF}}	= \\ 
\left(\begin{array}{rrrrr}
	-0.65 & 0 & 0 & 0.01+0.11i & 0 \\
	0 & -0.62 & 0 & 0 & -0.01-0.11i\\
	0 & 0 & -0.78 & 0 & 0 \\
	0.01-0.11i & 0 & 0 & -0.62 & 0 \\
	0 & -0.01+0.11i & 0 & 0 & -0.65
\end{array}\right)
	\end{aligned}
\end{equation}

\bibliography{Ref}

\end{document}